\documentclass{aa} 
\usepackage{epsfig,latexsym,longtable,lscape,supertabular,natbib} 
\bibpunct{(}{)}{;}{a}{}{,}

\def\e20{$\times 10^{20}$} 
\def\ergsec{erg s$^{-1}$}

\def\today{\number\day -\number\month -\number\year} 
\def\chandra{{\it Chandra}} 
 
\def\xmm{XMM-Newton} 
\def\ii{\,{\sc i}\,{\sc i}\,} 
\def\hi{H\,{\sc i}} 
\def\hii{H\,{\sc i}{\sc i}} 
\def\chandra{{\it Chandra}}

\begin{document} 

\pagenumbering{arabic} 
\title{Stephan's Quintet with XMM-Newton} 
\author{ 
G. Trinchieri$^1$, J. Sulentic$^2$,  W. Pietsch$^3$ and  D. Breitschwerdt$^4$ } 
\institute{ 
Osservatorio Astronomico di Brera, via Brera 28, 20121 
Milano Italy 
\and 
University of Alabama, Tuscaloosa, AL 35487, USA 
\and 
Max-Planck-Institut f\"ur extraterrestrische Physik, 
		Giessenbachstra\ss e, D-85740 Garching 
		Germany
\and
Institut f\"ur Astronomie, Universit\"at Wien,
        T\"urkenschanzstr. 17, A-1180 Wien, Austria} 
\offprints{G.~Trinchieri} 
\mail{ginevra@brera.mi.astro.it} 

\date{Draft: \today}

\abstract{The prototype compact group known as Stephan's Quintet (SQ)
was observed with XMM-Newton in order to complement the excellent
resolution of \chandra\ with high sensitivity to extended emission.  SQ
is a dynamic environment whose main effect, at both X-ray and optical
wavelengths, appears to be ISM stripping. This is
manifested by: 1) secular evolution of morphological types towards earlier
types and 2) growth of diffuse emission.  Virtually all cold, warm and
hot gas in SQ is found outside of the member galaxies. XMM-Newton
offers the opportunity to study the hot gas with unprecedented
sensitivity. We find two main components:  1) extended high surface
brightness emission from shocked gas associated with an ongoing
collision and 2) even more extended and unrelaxed diffuse emission that
follows the stripped stellar envelope of the group.
 } 

\authorrunning{Trinchieri et al.} 
\titlerunning{SQ with XMM-Newton} 
\maketitle 

\section{Introduction} 

Multi-wavelength data continue to accumulate for Stephan's Quintet (SQ)
the most studied of the compact groups of galaxies. A framework for
these observations has developed that may be quite generally relevant
to the compact group phenomenon. Groups like SQ represent the densest
galaxy aggregates in the non-clustered Universe. If SQ is representative,
then compact groups evolve and grow through a process of sequential
harassment by, and acquisition of, infalling intruder galaxies. It is
not clear what stimulates the infall because the observed baryonic mass
in SQ is about an order of magnitude too small to account for it.  Two
intruder galaxies have been identified in SQ \citep{Molesetal}. 
As summarized in
\cite{Trinchierietal2003}, the older, and
likely captured, intruder (NGC~7320c) has left evidence of two passages
in the form of parallel tidal tails. NGC~7319 was most likely the victim
of  the last passage having lost most of its ISM.  
From the evidence of an ``UV loop" in the new GALEX images of this
system,  \cite{Xuetal2005} suggest that only the
older tail  results from the passage of NGC~7320c, the younger one
being caused by a close encounter with NGC~7318a.  
In either case, the Seyfert 2 nucleus in NGC~7319 
may represent another manifestation of the encounter. 
Both it and NGC~7320c are examples of galaxies in transition from later (S) to
earlier (E/S0) type. One or both of the members of SQ with elliptical
morphology (NGC~7317 or 7318a) may have been stripped in earlier
passages of the old intruder.  The ``new'' intruder (NGC~7318b) recently
penetrated SQ at unusually high velocity ($\Delta$V$\sim$1000km/s) which
suggests that it is not bound to the system.
The strongest signs of current activity are related to this
latter encounter.

The results of the active dynamical history in SQ yield multifold 
observational manifestations of the interaction process: 1) tidal 
deformations, 2) ISM stripping, 3) star formation suppression within 
the galaxies, 4) star formation sites in the stripped debris, 5) 
morphological transformations, 6) creation of a common halo of stripped 
stars, 7) shocks and, possibly, AGN ignition.  It is widely assumed 
that interactions were more common in the early Universe and this makes 
detailed study of groups like SQ important beyond their benefit as 
laboratories for studying interaction physics. The signatures of the 
above interaction events have now been observed in all windows of the 
electromagnetic spectrum except the gamma ray.
SQ might therefore become a reference 
for extragalactic observers.

\begin{figure*}
\epsfig{figure=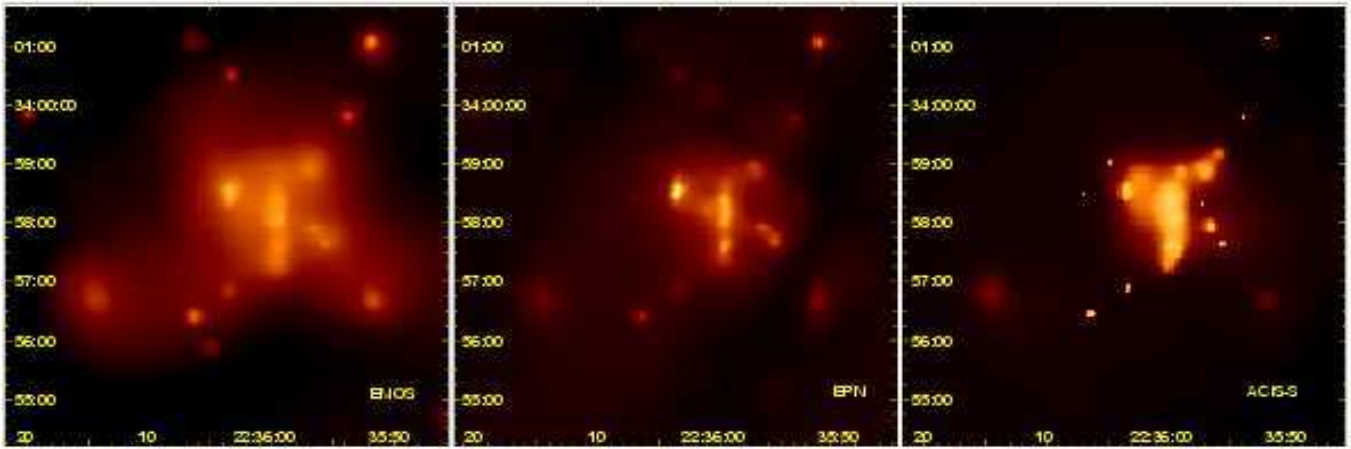,width=18cm,clip=}
\caption{Comparison of
XMM-Newton EPIC-MOS and EPIC-pn images with $Chandra$-ACIS
in the 0.3-3 keV band.  MOS1 and MOS2 are merged.  Adaptive
smoothing is applied using $csmooth$ in the Ciao package
(see {\tt http://asc.harvard.edu}). XMM images show lower resolution 
but enhanced sensitivity to extended emission.} \label{xmm+chandra}
\end{figure*} 

Recent \chandra\ observations of SQ \citep{Trinchierietal2003}
generally support the above evolutionary  picture and have shown
quite distinctly the effects of interaction between SQ  and its
intruders. A large scale shock ($\sim$40kpc at 85 Mpc) seen in both
ROSAT-HRI and \chandra\ images as a prominent narrow NS feature embedded
in more complex and extended diffuse emission is further confirmed by
these XMM-Newton observations (Fig.~\ref{xmm+chandra}).  The NS
structure was found to be  distinctly clumpy and sharply bounded on the
W side.  Spatial coincidence of the X-ray shock is found with both
radio continuum \citep{Williamsetal2002} and forbidden [N\ii] emission
\citep{Sulenticetal2001}
while little correspondence is seen between the
X-ray and optical broad band images.  The principal X-ray features are
best explained as manifestations of the shock produced by collision of
the new intruder with the debris field produced by passages of the old
intruder.   However, if the stripped ISM of NGC~7319 was largely displaced towards
the east, in the direction of NGC~7320c, as suggested by the observed
\hi\ 
distribution \citep{Williamsetal2002},  then currently shocked debris field must
be the product of an earlier encounter between the old intruder and,
perhaps, NGC~7318a.  The NS X-ray feature was interpreted \citep{Trinchierietal2003} as a
bow shock propagating through the pre-existing debris field
and heating it
to a temperature of $\sim 0.5$ keV. 
The low temperature of the post-shock is
a problem unless we postulate: a) an oblique shock and/or b) a weak shock
where the upstream medium is hot and has a sizable counter-pressure.  As
discussed in \cite{Trinchierietal2003}, a
standard weak shock is unlikely given the high Mach number of the inflowing
gas in the frame of reference of the new intruder.  In order to reduce
the post-shock temperature to a value consistent with that derived from
X-ray spectral fitting, the gas would have to enter the shock with
respect to the shock surface at an angle of $30^\circ$ or less. This will
only happen at the wings of the bow-shock. 

In \cite{Trinchierietal2003} we estimated the cooling time from X-ray spectral fitting and
a standard cooling function for a low metallicity plasma in collisional
ionization equilibrium to be roughly $4 \times 10^8$ yr for a mean density
of $n_X \sim 2.7 \times 10^{-2} \, {\rm cm}^{-3}$.  Note that in the
surrounding upstream medium the density can be lower by a factor of
$5 - 10$, so that cooling can be completely negligible, but that denser regions in
the shock have higher densities and correspondingly shorter cooling times. 

\begin{table} 
\caption[]{Log of the \xmm\ observations for Stephan's Quintet.} 
\label{log} 

\begin{tabular}{lll} 
\hline 
\hline 
Instrument & \multicolumn{2}{c}{Livetime (sec)}  \\ 
& Total & Used \\ 
EPIC-pn & 37088 & 31531 \\ 
EPIC-MOS1 &44845 & 36775\\ 
EPIC-MOS2 &44833 & 36766  \\ 
\hline \hline 
\end{tabular} 
\end{table} 

This interpretation is attractive
because it would also better explain evidence that the new intruder
shows considerable tidal distortion. In this case {\it in flagrante
delicto} implies an intruder crossing time of t$_c \sim 10^8$ years.

\begin{figure*} 
\epsfig{figure=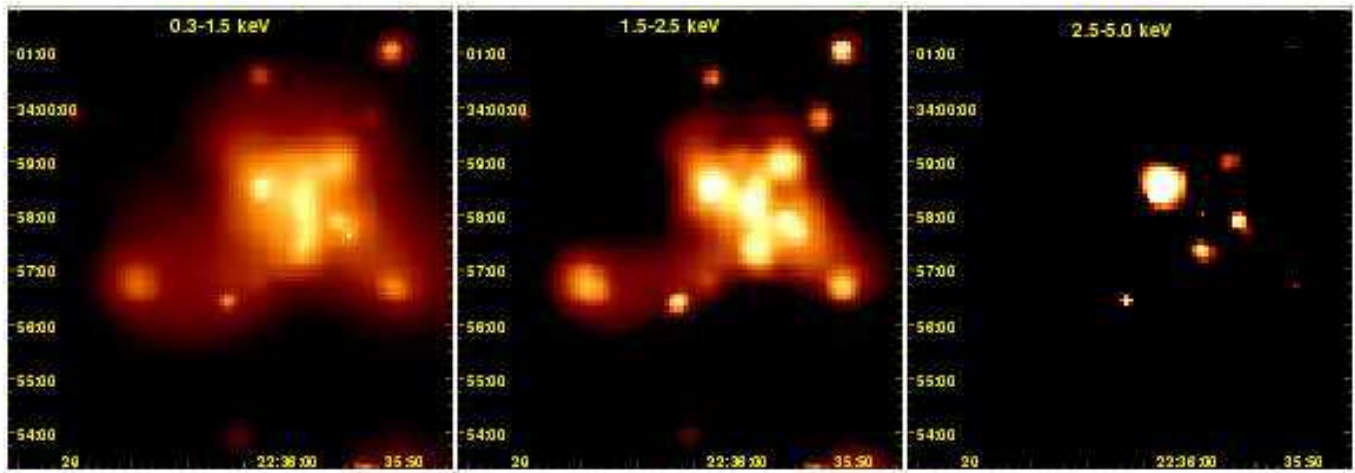,width=18cm,clip=} 
\caption{Comparison of smoothed EPIC-MOS images in the 0.3-1.5 keV (LEFT), 
1.5-2.5 keV (CENTER) and 2.5-5.0 keV (RIGHT) bands. Extensive diffuse 
emission is evident in the softest image.} 
\label{xmm3bands} 
\end{figure*} 

\section{XMM-Newton Data Analysis} 

SQ was observed for $\sim$40 ks with EPIC-MOS and EPIC-pn during 
revolution 366 of XMM-Newton. The event files were reprocessed with the 
$xmmsas-5.4.1$ release and further cleaned of residual high background 
due to flaring activity.  We used the EPIC-pn light curve for 
processing data from all instruments. Only photons with FLAG=0 and 
PATTERN$\le$4 (for EPIC-pn), or PATTERN$\le$12 (EPIC-MOS), were used in 
the analysis. The final exposure times for the accepted events are 
listed in Table~\ref{log}. Data analysis employed both the 
$xmmsas$/ftools and $ciao$/funtools tasks with heavy reliance on ds9 
and corollary software. 

\begin{figure*} 
\resizebox{18cm}{!}{ 
\epsfig{figure=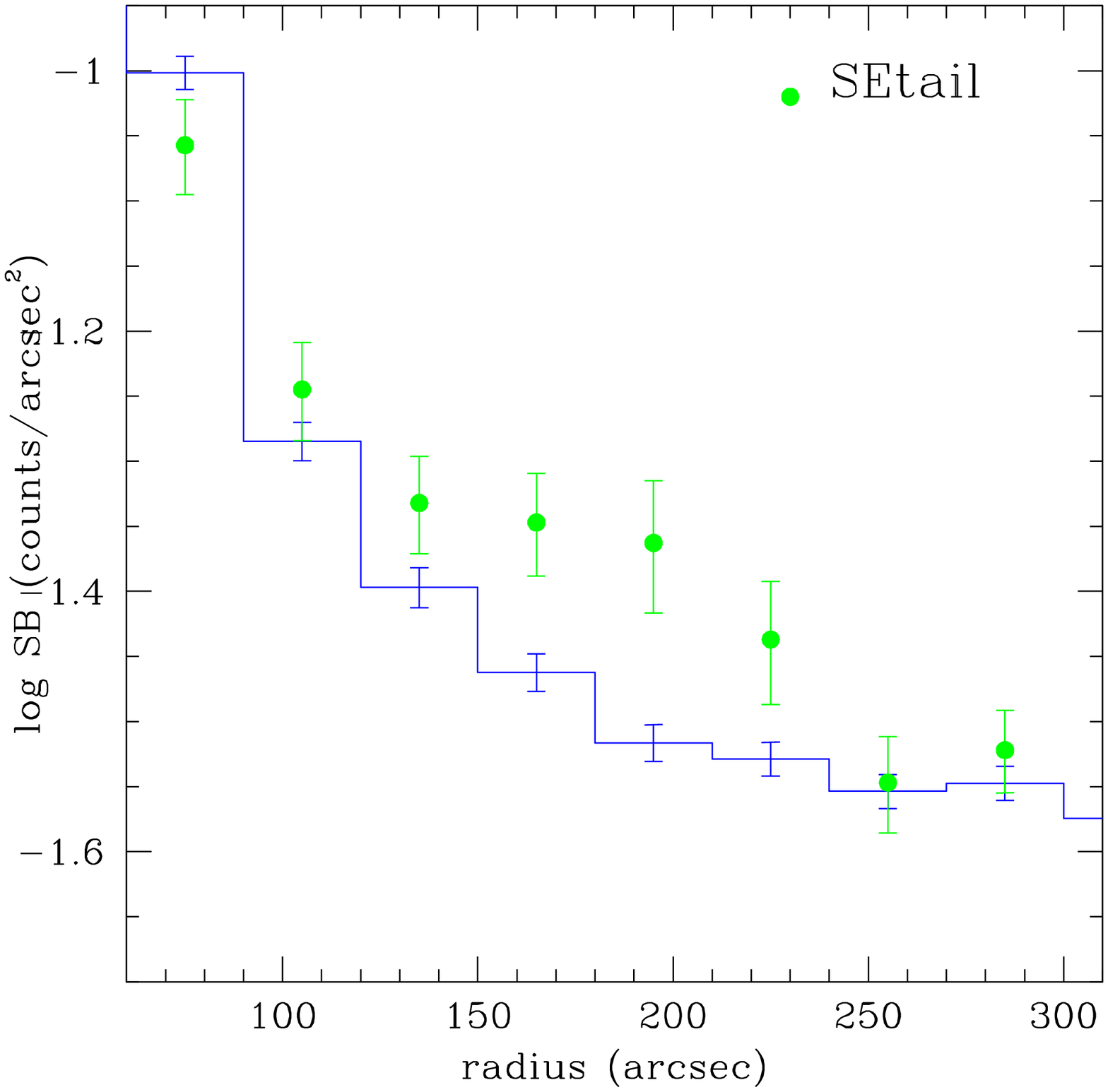,width=18cm,clip=} 
\epsfig{figure=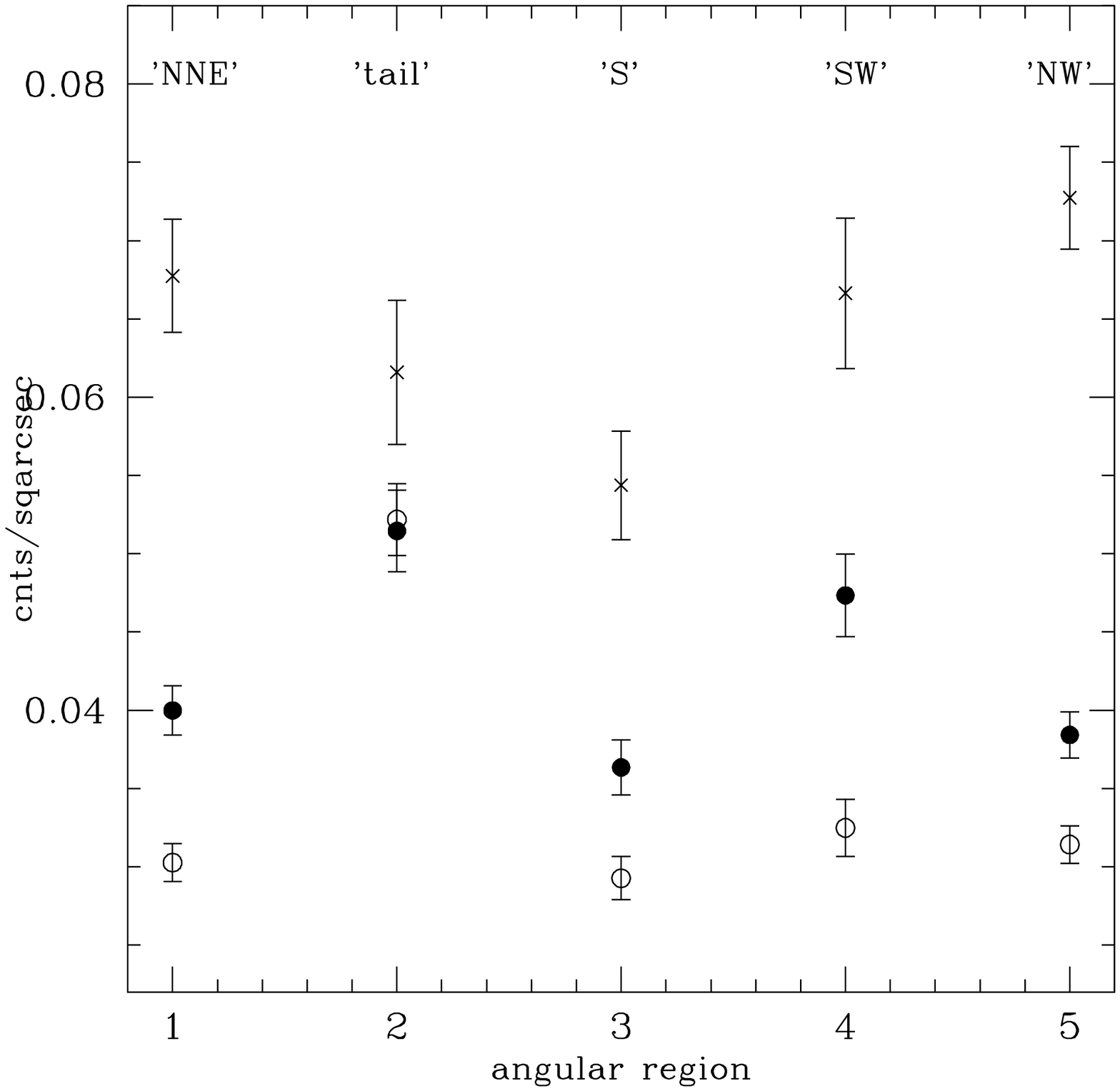,width=18cm,clip=} 
} 
\caption{LEFT: 
The radial distribution of X-ray emission in the  azimuthal ranges
95$\degr$-140$\degr$ (filled dots, SEtail) and complementary 140$\degr$-95$\degr$
(histogram), 
obtained from the combined
EPIC-MOS data (0.3-2.5 keV) that are free of CCD gaps. The former shows
the excess diffuse emission in the direction of the tidal tails.
RIGHT: Azimuthal distribution of the emission at different radial
distances from the middle blob in the NS feature: Crosses:
r=$75''-114''$; Filled dots: r=$114''-180''$; empty
dots: r=$180''-240''$.  The sources associated with the Seyfert NGC~7319
and with NGC~7317 are masked out.   Azimuthal angles are:
NNE=0$\degr$-95$\degr$; tail=95$\degr$-140$\degr$;
S=140$\degr$-210$\degr$; SW=210$\degr$-255$\degr$;
NW=255$\degr$-360$\degr$.
} 
\label{radprof} 
\end{figure*} 

\subsection{X-ray Images: characterization 
of the diffuse emission}

In an effort to enhance the signal-to-noise  we merged the central CCD
frames of the two EPIC-MOS observations and smoothed the result with an
adaptive algorithm ($ciao-csmooth$). EPIC-pn data are kept separate
because of the different patterns of CCD gaps in EPIC-pn and EPIC-MOS
instruments.  EPIC-pn, EPIC-MOS and \chandra\ \citep{Trinchierietal2003} images (0.3-3.0
keV) are compared in Fig.~\ref{xmm+chandra}, which illustrates the
complimentary power of data from the two satellites.  While the
\chandra\ image clearly shows more discrete features, the large
collecting area of  XMM-Newton facilitates detection of diffuse X-ray
emission components.  A comparison of EPIC-MOS emission in different
energy bands is shown in Fig.~\ref{xmm3bands}.  Both figures reveal the
complexity of the emission in SQ with a multitude of both compact and
extended features centered on the NS oriented shock, with different
relative strengths as a function of the energy band considered (see
later).  A previously undetected component is visible towards the SE as
an extension that connects the low surface brightness emission at the
center of the group with an enhancement that was marginally detected in
the \chandra\ data \citep{Trinchierietal2003}. This new component, called TAIL for easy
reference in the text, is faint even in XMM-Newton images making a full
characterization of its properties very difficult.

\begin{figure*} 
\epsfig{figure=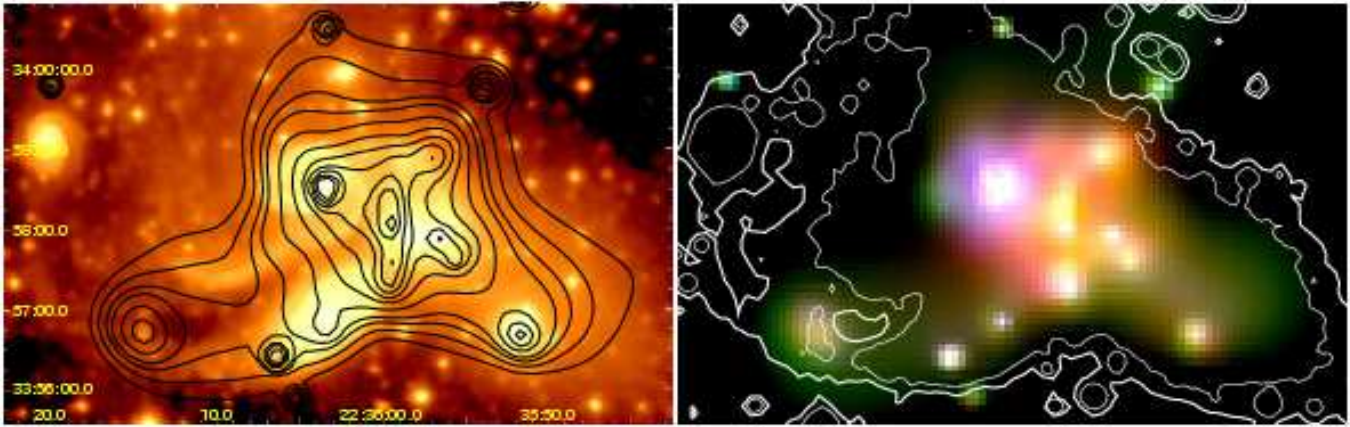,width=18cm,clip=}
\caption{
(LEFT) Merged EPIC-MOS contours superimposed on a deep
R-band image taken at the 2.5m INT (courtesy of  Carlo Gutierrez).  The
0.3-3.0 keV band  X-ray data were smoothed with an adaptive algorithm before contouring.
Outer contours chosen to show maximum extent of X-ray
emission.  (RIGHT) Optical R band contours superimposed on
a  ``true'' color X-ray image produced from 3 different EPIC-MOS energy
bands:  red:0.3-1.5 keV; green=1.5-2.5 keV; blue=2.5-6.0 keV. Contours 
chosen to show maximum extent of the diffuse halo.}
\label{color} 
\end{figure*} 

The more sensitive XMM-Newton data allow a better definition of low
surface brightness emission in SQ and we have tried to determine its
extent by means of radial profiles of the X-ray emission.  Regions of
CCD gaps, bad columns and bad pixels identified on the exposure maps
are masked and all detected sources outside a radius of $\sim 3'$ from
the center of SQ are excluded. Emission above $\sim 2.5-3.0$ keV is
associated with the nucleus of NGC~7319 and a few much fainter
individual sources.  Diffuse emission is detected out to a maximum
extent of r$\sim 4'$  at energies below $\sim 2.5$ keV.  However, the
emission becomes clearly nonuniform with increasing distance from the
central shock region, so that the extent depends on the assumed
direction.  Figure \ref{radprof} shows both the clear excess in the
TAIL region identified above (left panel) and the azimuthal dependence
of the extent of the emission (right panel), towards the SE (TAIL) and
the SW (towards NGC~7317) at an average radius of $2'-3'$.  The smaller
extent towards S could reflect the presence of additional absorption,
due in part also to the foreground NGC~7320 (however, the same should
be true for TAIL).  The EPIC-MOS CCD gaps make any characterization of 
any diffuse structure outside r$\sim 4'$ difficult to quantify properly.

\begin{figure} 
\epsfig{figure=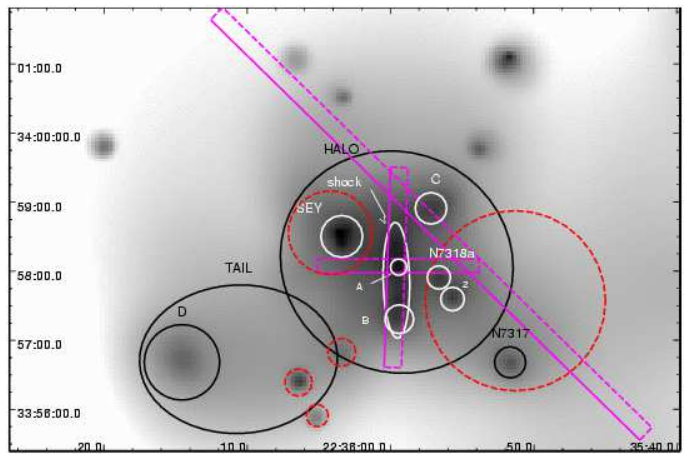,width=9cm,clip=} 
\caption{ 
Identification of source regions used in the production of pseudo-spectra
(Fig.~\ref{colorb}). Unlabeled red regions within a marked area were 
excluded from the spectral analysis of that region. 
Also shown are the regions used to  produce the cuts shown in Fig.~\ref{xhaprof}
and Fig.~\ref{xmmred}
} 
\label{whichreg} 
\end{figure}

\begin{figure} 
\epsfig{figure=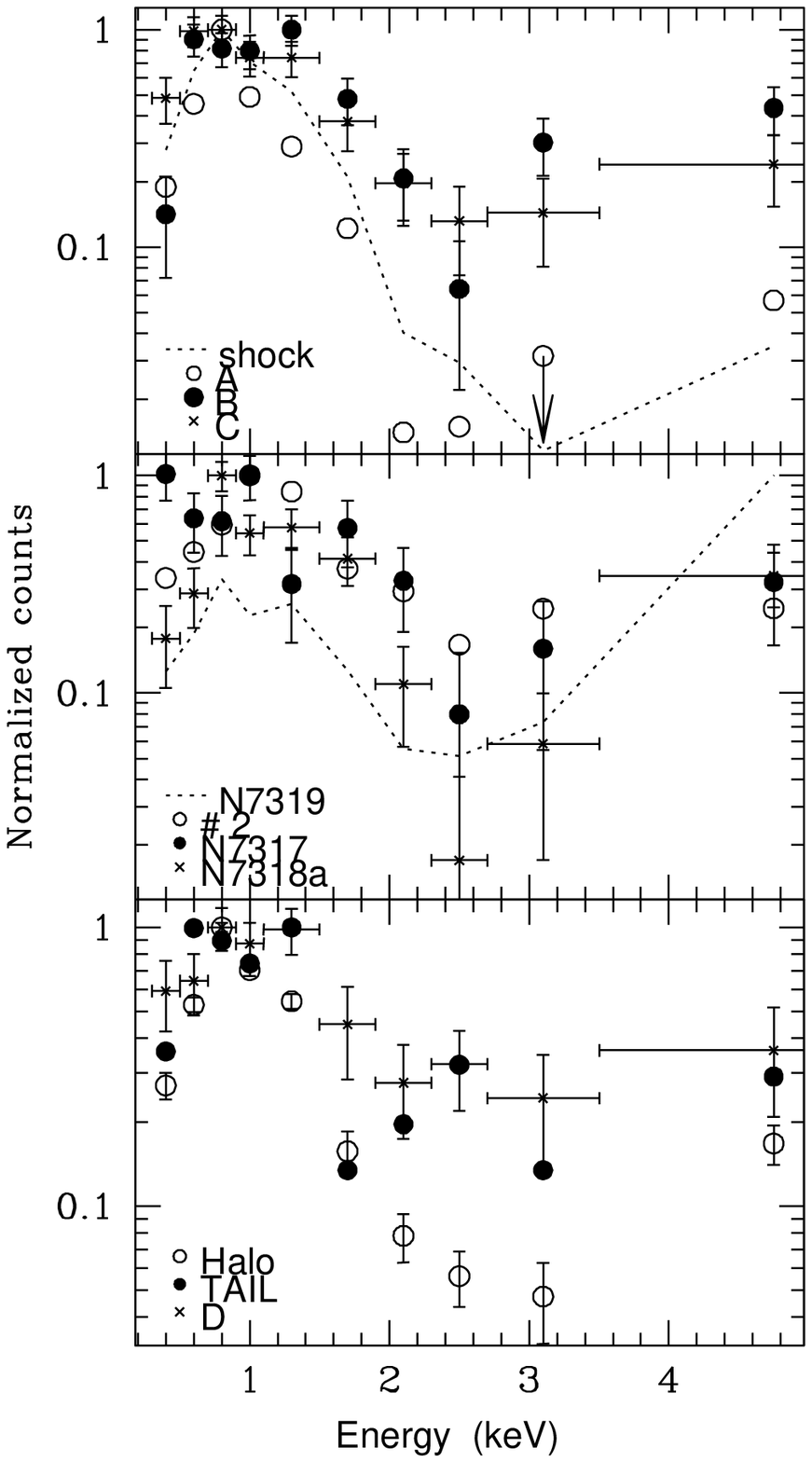,height=14.7cm,clip=} 
\caption{ 
Photon energy distributions (pseudo-spectra) for regions highlighted 
in Fig.~\ref{whichreg}. No correction for the spectral response is applied,
so plotted distributions are not representative of the real spectral shapes. 
Counts are binned all in the same energy ranges (shown by horizontal
bars) and normalized to the highest value so that relative shapes are 
immediately comparable to one another. 
Uncertainties are given only for selected 
regions to preserve clarity in the plots. Photons from smaller 
regions included in larger ones are not included in pseudo-spectra
for these latter. TOP: Regions in or near the central shock region.
MIDDLE: Sources associated with SQ galaxies plus the 
single detected giant emission region SW of NGC~7318a (\#
2). Because of 
the much stronger emission from the Seyfert nucleus of NGC~7319 in the 
hard band, that determines the relative normalization of the
pseudo-spectrum, all points are systematically below other objects. 
BOTTOM: Regions of diffuse emission 
outside the central shock region. 
} 
\label{colorb} 
\end{figure}

In order to better highlight different spectral signatures in the 
complex emission from SQ we have generated a color image combining data 
in three broad bands: 0.3-1.5, 1.5-2.5 and 2.5-7.0 keV (displayed as 
red, green, and  blue respectively).  The resulting image, shown in 
Fig.~\ref{color} shows a number of localized regions with different 
colors that imply differences in photon energy distribution. Many of 
these regions lack sufficient photons for a detailed spectral analysis 
but we can make a rough comparison of the relative photon distributions 
for some of the regions that are illustrated in Fig.~\ref{whichreg}. 
A more detailed analysis is given for regions with high enough 
photon statistics (see \S~\ref{spectralsection}).  Constant broad 
energy bins are used for constructing all of the ``pseudo-spectra'' 
displayed in Fig.~\ref{colorb}. These energy distributions have not 
been corrected for instrumental response. They were extracted from 
relatively small areas of the central CCD in EPIC-MOS so we do not 
expect variations in the response matrix to affect this qualitative 
comparison between different sources and regions, but cannot be used to derive
spectral shapes. 

The pseudo-spectra show significant differences that suggest 
these sources are unlikely to be fit with the same 
spectral model. We comment on the most striking:

\noindent 1) Regions  B and C (top panel) show similar distributions especially compared 
with region A.  The average pseudo-spectrum of the shock region is
intermediate between A and B distributions.  \chandra\ data suggest that a 
compact source (\chandra\ \#5 \citealt{Trinchierietal2003}) is embedded in condensation C (but 
not at its center).  With the current astrometry and published
positions of the H$\alpha$ knots \citep[e.g.][]{Xuetal2003},  the \chandra\ position does not suggest 
coincidence with any obvious optical feature in the new intruder, so it 
might be an unrelated background source. A region of intense star 
formation \citep[STARBURST A in][]{Xuetal1999} and associated \hi\ and CO emission
lie between the N end of the shock and source C.

\noindent 2) The  sources in the middle panel are associated with  
SQ galaxies and show similar pseudo-spectra. NGC~7319 shows a strong hard 
photon excess which is likely related to an obscured active nucleus. 
Significantly different line-of-sight absorbing column densities might 
explain the large difference between the lowest energy points of NGC~7318a and 
NGC~7317. This is consistent with the lack of any radio line or continuum
detections near to NGC~7317. Unresolved radio continuum emission is detected 
from NGC~7318a \citep{Williamsetal2002, Xuetal2003}. Source \# 2 is also 
included in this panel.  Thus is the only one of four very large (D$\sim$400 pc) 
H$\alpha$ emission condensations detected as a discrete X-ray source
\citep{Trinchierietal2003} and belongs to NGC~7318b \citep{Sulenticetal2001}. 

\noindent 3) Diffuse emission regions outside the central shock region (bottom
panel) do 
not show identical spectra: TAIL and D are consistent with each other 
but not with HALO.  The latter feature is more reminiscent of the 
spectral distribution of the shock front and may be evidence for
a large scale signature of the shocked new intruder disk.
This will complicate any inferences about the extent and integrated 
properties of diffuse X-ray component that existed before the 
arrival of the new intruder.  It is unclear if D is a 
condensation near the end of this tail or an unrelated background
source: there is evidence of an extended object at its center
(redshift and nature unknown at the present time), but at the same
location there is a clear deficit in the \hi\ distribution that, if
foreground, would represent a smaller absorbing column with consequent
local enhancement of the emission. The location of D also coincides with
the ``end" of the \hi\ in Arc-S \citep{Williamsetal2002} and could also
represent the edge of the expansion of the hot gas. 
The similarity of the spectral distributions for
D and TAIL might favor the interpretation of D as a feature associated
to the more diffuse component in TAIL. 

\begin{table*}
\caption[]{Net counts in the 0.3-3 keV energy band in the PN and m1+m2
detector for selected regions, also shown in Fig.~\ref{whichreg}.  
The background is taken from an ellipse
to the E, in a neighbouring CCD in EPIC-pn (see text).  Fluxes and luminosities 
(0.3-3.0 keV band) are computed assuming
a conversion factor of 4$\times 10^{-17}$, for EPIC-pn counts, corresponding to a
line-of-sight value for absorption and a power law with $\Gamma=1.7$, and a distance of
85 Mpc.
}
\label{sources}
\begin{tabular}{lllcrllll}
\hline
\hline

Src.&Chandra$^1$ &\multicolumn{1}{c}{Position}& radius& PN counts & MOS counts   & flux& luminosity
\\
\#&name&\multicolumn{1}{c}{(J2000)}&$^{''}$&and error&and error&cgs&cgs\\
\\	
1&1  &22:35:53.96,+33:59:45.9 &11.25 &   73.2$\pm$10.2 & 58.7$\pm$ 8.9&2.9$\times 10^{-15}$&  2.5$\times 10^{39}$\\
2 &2  &22:35:55.68 ,+33:57:35.9 &10.25&  186.5$\pm$14.1 &   97.7$\pm$10.3&7,5$\times 10^{-15}$&  6.5$\times 10^{39}$\\
NGC7318a &3  &22:35:56.47 ,+33:57:54.7 &10.15 &  218.9$\pm$15.2&  129.0$\pm$11.7&8.8$\times 10^{-15}$&  7.7$\times 10^{39}$\\
NGC7317& $-^2$&22:35:51.66 ,+33:56:40.8&19.60& 101.9$\pm$11.1  &  90.9$\pm$10.3&4.1$\times 10^{-15}$&3.1$\times 10^{39}$ \\
C &5&22:35:57.18,+33:58:54.7& 13.70 & 337.7$\pm$19.0 & 217.4$\pm$14.3&1.4$\times 10^{-14}$&  1.2$\times 10^{40}$ \\
D&  $-^2$ &22:36:14.55 ,+33:56:41.0 &13.60  &  285.3$\pm$21.2 &204.3$\pm$17.8&1.1$\times 10^{-14}$&  1.0$\times 10^{40}$\\
NGC7320nuc&6-7  &22:36:03.41 ,+33:56:49.7 &12.10  &   88.4$\pm$10.3&   49.9$\pm$ 7.8&3.5$\times 10^{-15}$&  6.0$\times 10^{37\dagger}$\\
NGC7320SE& 11  &22:36:06.42 ,+33:56:23.5 &11.80&  118.7$\pm$11.7&    83.4$\pm$ 9.8 &4.8$\times 10^{-15}$&
  8.3$\times 10^{37\dagger}$\\
\hline \hline
\end{tabular}

$^\dagger$ In NGC 7320, so we assume a distance of 12 Mpc \\
$^1$ Positional coincidence.  However the sizes of the XMM-Newton regions  might include a larger
fraction of extended emission \\
$^2$ Not detected by Chandra \\
\end{table*}
\subsection{Characterization of individual features} 
\label{spectralsection}

The previous images and pseudo-spectra enable us to identify emission
regions where it  is possible to derive both the contribution of
individual components  (see Table~\ref{sources}) and where the photon
energy distribution allows a more detailed spectral analysis.
Table~\ref{sources} lists detected counts, fluxes and luminosities for 
individual sources detected and already presented in Fig.~\ref{whichreg}
for which a more detailed spectral
analysis is not possible.   Regions with stronger emission 
are instead considered for more detailed spectral modeling, as summarized in 
Table~\ref{spectab}.  Source shapes
and sizes were adjusted to avoid regions of non-uniform illumination
(e.g. CCD gaps), but were kept as large as possible (compatible with
colors and morphologies implying coherent components, as suggested by
both our X-ray color image and Fig.~\ref{colorb}) in order to improve
the signal-to-noise in the derived spectra.  Photons were binned to
achieve a minimum signal-to-noise  $\ge 2-3 $ (depending on the
resulting number of bins) in each energy bin after background
subtraction. We generated $arf$ and $rmf$ files, using the most recent
calibrations.  We used PATTERN$\le$12 for EPIC-MOS[1-2] (singles,
doubles, triples and quadruples) and PATTERN=0 for EPIC-pn (singles
only), since we are mainly interested in softer energies where the
``single'' percentage was highest.  We also considered ``doubles'' in
the Seyfert data that contains many photons at higher energies in order
to increase the signal-to-noise.  Results of spectral modeling are
summarized in Table~\ref{spectab}. 
The data do not provide meaningful constraints on the choice of spectral
model, with several different assumptions yielding equally good results. 
In particular, we have made two different
assumptions for the abundance parameter, fixing it at 30\% of cosmic
(consistent with the choice in \citealt{Trinchierietal2003}), and at
100\%. The former value gives in general better fits, i.e., lower $\chi^2$
values: an additional component is needed to
obtain a similar $\chi^2$ under the assumption of 100\% cosmic; 
we have chosen a power law with fixed slope at $\Gamma=1.7$, but other
choices give equal results.  The spectral parameters are generally consistent
in the two
models.  In fact the error regions are ill-defined in several
cases, due to apparent non-monotonicity in the $\chi^2$
space.  As discussed later \citep[see also][]{Trinchierietal2003}, the
spectral parameters should be regarded as only indicative, as all
regions are probably complex, and might contain different components at
different temperatures.   Moreover, 
plasma temperatures are derived from models
of optically thin emission, in which the plasma is assumed 
in collisional ionization
equilibrium.    In the presence of  non-equilibrium plasma, such as in
cases of shocks, these models would 
have the effect of ``artificially" requiring a multi-temperature fit.

\begin{figure*}
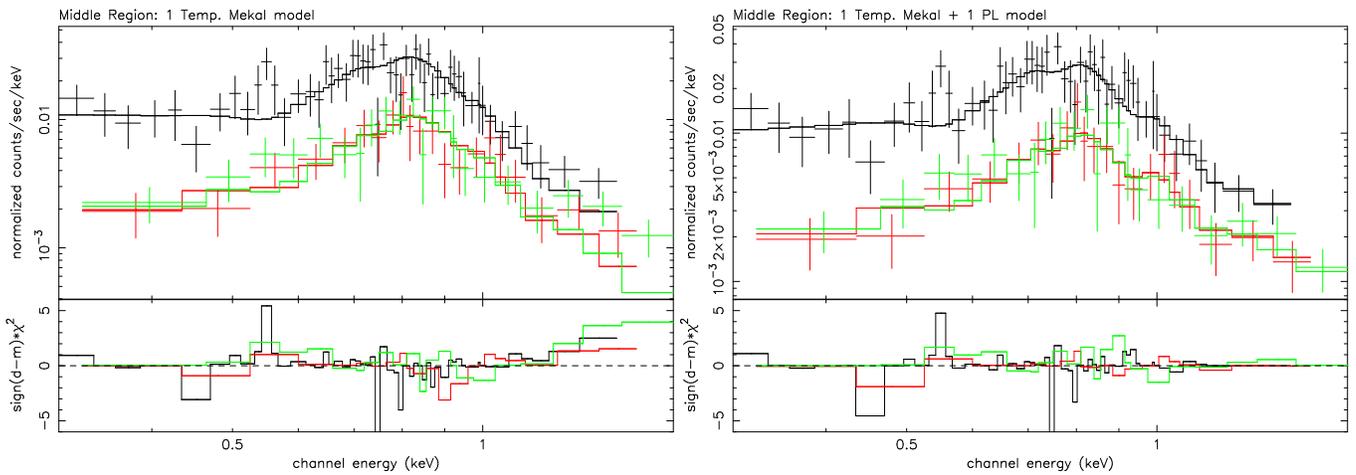
 
\resizebox{18cm}{!}{ 
\epsfig{file=onet,width=9cm,angle=-90} 
\epsfig{file=twocomp,width=9cm,angle=-90} 
} 
\caption{Spectral energy distribution for the central part of the shock 
including condensation A. 
Left: a single temperature plasma model is used to fit the data (see 
Table~\ref{spectab}).  Right: a power law is added to the 
model. The best fit temperature and galactic absorptions vary slightly 
from the original fit 
but are consistent within the errors. 
} 
\label{mb} 
\end{figure*} 

The background is chosen from a region to the E in the adjacent
chip for EPIC-pn (but within the same CCD for EPIC-MOS) where we see
and expect no emission from SQ.  We have also considered other regions
to the N and W (also in adjacent EPIC-pn CCDs) but found no significant
variations.  
Table~\ref{spectab} also lists the luminosities of the different
regions for the different best fit parameters.  In a few cases, where
the best fit absorption is different, the softer band luminosity could
be significantly different in the 30\% and 100\% abundance assumption.

\begin{table*}
\caption[]{
Spectral results for the different regions defined in the
text. The thermal model is a mekal with the stated percentage of 
cosmic abundance. 
}
\begin{tabular}{lllrrlllllll}
\hline
\hline
\multicolumn{5}{l}{Region (cf. Fig.~\ref{colorb})} &
\multicolumn{5}{l}{Counts and errors} \\ 
 Z   & \multicolumn{1}{c}{ N$_h$ }& kT$_1$& kT$_2$&
N$_h$ & $\Gamma$ &$\chi^2(DOF)$ &$\chi^2_{\nu}$ & L(Total) &
L(Total) &  L(kT$_1$) & L(kT$_2$) \\
&&&&&&&&[0.5-2.0] & [2.0-10.0] & [0.5-2.0]& [0.5-2.0]  \\
\% & cm$^{-2}$ & keV& keV &  cm$^{-2}$&  & & & erg s$^{-1}$ &
erg s$^{-1}$ & erg s$^{-1}$ & erg s$^{-1}$ \\
\hline \hline	     
\multicolumn{5}{l}{SHOCK (includes A, B) }& \multicolumn{6}{l}{PN: 1558$\pm$40
MOS1:  585$\pm$25 MOS2: 613$\pm$25 }\\
30 & 0.28  &0.25
&0.84&--&--&195.6 (195)& 1.0 &2.4$\times10^{41}$&4.1$\times10^{39}$ &
 1.9$\times10^{41}$& 4.7$\times10^{40}$  \\
100 & 0.36& 0.23 & 0.98 &--&-- & 214.9 (195)& 1.1 & 3.5$\times10^{41}$&
4.1$\times10^{39}$ & 3.2$\times10^{41}$& 3.9$\times10^{40}$\\
100 & 0.19&0.27 & 0.81 &--& 1.7$^*$ & 193.6 (194) & 1.0& 1.4$\times10^{41}$ &
2.5$\times10^{40}$& 1.4$\times10^{41}$ & 2.1$\times10^{40}$ \\
\hline
\multicolumn{5}{l}{~~~In shock: Southern Tip }&\multicolumn{6}{l}{PN: 337$\pm$19
MOS1: 110$\pm$11 MOS2: 99$\pm$10 } \\ 
30 & 0.49 & 0.22 & -- &--&& 46.5 (42)&1.11 & 1.8$\times10^{41}$
&4.9$\times10^{37}$ \\
100 &0.49 & 0.22 &  -- &--&& 48.3 (42)&1.15 & 1.7$\times10^{41}$
&5.1$\times10^{37}$ \\
\hline
\multicolumn{5}{l}{~~~In shock: Middle Region ($\sim$A)}
&\multicolumn{6}{l}{PN: 470$\pm$21 MOS1: 177$\pm$13 MOS2: 207$\pm$15} \\
30& 0.03$^{*}$ & 0.59 &-- &-- &&87.6 (93)&0.95 &2.0$\times10^{40}$
 &5.6$\times10^{38}$ \\
 30 & 0.07 & 0.50 & --&-- &1.7$^*$& 65  (92)&0.7&2.5$\times10^{40}$
 &1.5$\times10^{40}$ & 1.6$\times10^{40}$ \\
 100 & 0.03$^{*}$ & 0.59 & --&-- &&122 (93)&1.31  &1.9$\times10^{40}$
 &3.8$\times10^{38}$ \\
 100 & 0.04 & 0.50 &-- & --&1.7$^*$& 65  (92)&0.7&2.2$\times10^{40}$
 &1.4$\times10^{40}$ &1.5$\times10^{40}$  \\
\hline
\multicolumn{5}{l}{~~~In shock: Northern  Tip }&\multicolumn{6}{l}{PN: 527$\pm$23
MOS1: 207$\pm$15  MOS2:  183$\pm$14}\\
30& 0.16& 0.30 & 1.06 & --&--&89.1 (72) & 1.24 &2.0$\times 10^{40}$&
1.9$\times10^{39}$ & 1.2$\times10^{40}$ & 8.2$\times 10^{39}$ \\
100 & 0.32& 0.24 & 1.27 &--&--& 98.2 (72)&1.36 & 9.8$\times 10^{40}$& 2.4
$\times 10^{39}$& 8.5$\times 10^{40}$& 1.3$\times 10^{40}$ \\
100 & 0.11& 0.32 & 0.93&--&1.7$^*$& 86.4 (71) &1.21 & 3.2$\times 10^{40}$ & 1.3
$\times 10^{40}$ & 1.8$\times 10^{40}$ & 6.5$\times 10^{39}$\\
\hline 
\multicolumn{5}{l}{HALO}&\multicolumn{6}{l}{PN: 2820$\pm$63 MOS1:  1184$\pm$42
MOS2:  1157$\pm$ 42} \\
30 & 0.12  &0.30 &0.77&--&1.58&216.6 (196) & 1.1 & 1.9$\times10^{41}$&
5.3$\times10^{40}$  & 9.5$\times10^{40}$& 7.1$\times10^{40}$\\
100  & 0.12  &0.29 & 0.76 &--& 2.20 & 219.2 (196) & 1.12 & 1.8$\times10^{41}$&
4.1$\times10^{40}$  & 8.6$\times10^{40}$& 5.4$\times10^{40}$ \\
\hline
\multicolumn{5}{l}{TAIL (Includes D)} &\multicolumn{6}{l}{PN: 441$\pm$33
MOS1: 227$\pm$23 MOS2: 210$\pm$23 } \\
30& 0.42& 0.16 & 1.50&--&--&33 (32) & 1.05 & 1.6$\times10^{41}$&
1.0$\times10^{40}$ & 1.4$\times10^{41}$& 2.5$\times10^{40}$  \\
100 & 0.72 & 0.1$^{*}$ & 1.44&--&--&35 (32) & 1.09 & 1.7$\times10^{42}$&
1.1$\times10^{40}$ & 1.6$\times10^{41}$& 3.4$\times10^{40}$   \\
\hline
\hline
\multicolumn{5}{l}{Seyfert galaxy$^{**}$}&\multicolumn{6}{l}{PN: 3310$\pm$57
MOS1: 993$\pm$32  MOS2: 1029$\pm$32 }\\
30 & 0.04 & 0.60 &  --& 45 & 1.33
&421  (401)& 1.05  &8.3$\times10^{41}$& 2.7$\times10^{42}$& 2.5
$\times10^{40}$& 2.6$\times10^{42**}$ \\
100 & 0.02 & 0.60 & -- &    48 & 1.48 &425 (401)& 1.06 & 1.1
$\times10^{42}$& 2.9$\times10^{42}$ & 2.0$\times10^{40}$&
2.9$\times10^{42**}$\\
\hline \hline
\end{tabular}

\noindent NOTES \\
$^*$ Fixed value of $\Gamma$, and  fixed lower limit for N$_H$ or kT  \\
$^{**}$ For the EPIC-PN data we have used both single and double events. 
Intrinsic luminosities 
refer to total (all three components), plasma (0.5-2.0 keV band)
and nuclear components (2-10 keV band), 
respectively. 
\label{spectab}
\end{table*}

\underline{SHOCK region}.  The shock region is a clumpy NS-oriented
elongated feature (Fig.~\ref{whichreg}). This structure is similar to the
one observed with \chandra. It includes condensations A and B but not
C.  The new X-ray data require a two temperature plasma,
at $\sim 0.3$ and $\sim
1$ keV with N$_H$ relatively high (3$\times$ the line-of-sight value).

Modeling the whole region is of limited significance
given the differences observed in the photon spectral distributions of
the sub-regions highlighted by the color image and the pseudo-spectra
(Fig.~\ref{color}-\ref{colorb}). 
Analysis of different condensations within the shock confirms their different 
spectral properties.  We have considered three separate
regions as indicated in Table~\ref{spectab} corresponding to the three main peaks
determined by \chandra\ and confirmed in XMM-Newton. The middle one is larger
than, but inclusive of, A. Assuming again plasma models, 
the middle and southern blob can be modeled with a single
temperature but with different values for kT (0.6 and 0.2 keV
respectively) and significantly different line-of-sight N$_H$. This may
reflect in part the larger \hi\ column \citep[$>$6$\times$10$^{20}$
atoms cm$^{-2}$;][]{Williamsetal2002} along
the line of sight due to the foreground galaxy NGC~7320 that partially covers the southern tip of the shock region.

A hard excess is visible in the data for region A (Fig.~\ref{mb}),
regardless of the abundance parameter, that
can be accounted for by a second component (any model will do), even
though the fit does not formally require a two component model
($\chi^2_{min} \sim 1$). No equivalent component is required in the
other regions (except for the 100\% abundance assumption, but see
above).  
A  two temperature model, with a relatively high amount
of absorbing material, is instead required for the northern tip. 

\underline{HALO region.}  All high surface brightness regions related 
to the shock, as well as discrete sources identified in
Fig.~\ref{whichreg}, 
were excluded from the photon distribution yielding our best attempt to
isolate a pure halo component.  A two temperature 
model could represent the data, 
as in the previous region, but with a marginally credible $\chi^2$=262 for 198 
Degrees of Freedom, and high temperature values of  kT$_1$=0.6 and kT$_2 =4.2 $
keV, with lower than galactic absorption 
N$_H$ = 0.04 (for 30\% cosmic abundance; at 100\% cosmic,  $\chi^2$=274,
for similar spectral parameters).  Addition of a third component brings a
significant improvement in the fit quality and 
the temperature values closer to
those observed in other regions (see Table~\ref{spectab}). 
In particular, the addition of a power law component gives a
$\Delta\chi^2$ of 45-55 (30-100\% abundance respectively) 
for 2 additional parameters. 

There is a likely
possibility of contamination from  discrete sources (one such was
detected by \chandra, \# 4 in \citealt{Trinchierietal2003}, but not by
XMM-Newton).  However contamination should affect all regions equally,
and we have subtracted all visible sources, so any contamination should
be of low level.  

\underline{TAIL region}. Counts from both source D and the 
connecting region were considered in order to improve the
signal-to-noise.   This 
combination is supported by the pseudo-spectra plot (Fig.~\ref{colorb}) that 
indicates a similar photon distribution for the two features. On the 
other hand it could be dangerous because we do not know if D belongs to SQ.   
Even taking this risk we still have a small number of photons and the 
resulting parameters are not well determined.  Although a
single temperature model could  be marginally  acceptable
($\chi^2_{min}$=1.4 for 34 DoF), the best fit
parameters (N$_H$ significantly smaller than the line-of-sight value,
kT$>2$ keV) are probably unreasonable for the system. 
A two component model gives a significantly better fit 
($\Delta\chi^2 >$ 14 for 2 additional parameters yielding  an $f-test$ 
probability P$>99.99$ that the additional model component give an improved 
fit).  Table~\ref{spectab} presents the results for a 2-plasma model 
fit.  Equivalent results are found with plasma+thermal or plasma+power law 
models.  

\begin{figure*}
\epsfig{figure=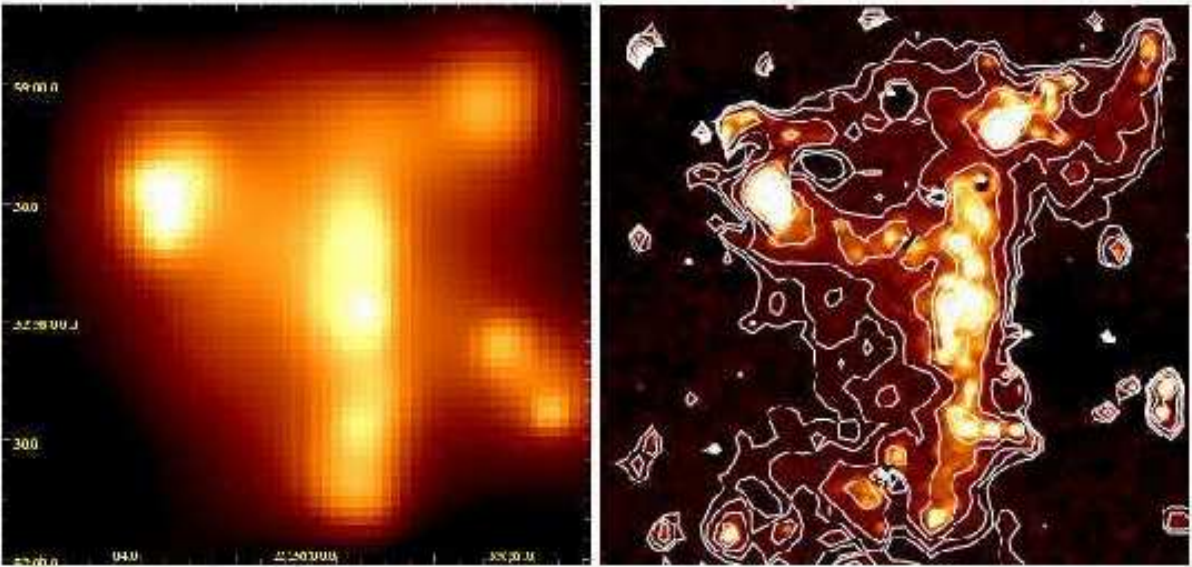,width=16cm,clip=}
\caption{Comparison between X-ray (smoothed EPIC-MOS data in 0.3-3.0 keV band,
LEFT) and H$\alpha$+[N\ii ] ($\sim 6600$ km s$^{-1}$, RIGHT)  emissions in SQ. 
The emission at the bottom of the right panel, to the 
SE of the bright NS feature, is contaminated by
[S\ii]$\lambda$6717,6731 emission in NGC~7320 that falls in
the filter passband.
\label{hasoXima}}
\end{figure*}

\underline{NGC~7319.} The spectrum of NGC~7319 is clearly complex and
requires several components as we already found in our \chandra\ analysis.
We now know that the discrete \chandra\ source \#9 located 8 arcsec S
of the nucleus is a z=2.2 QSO \citep{Galianni}   however it contributes
only about 5\% of the photons near to NGC~7319.  The  Seyfert nucleus was
fit with a double power-law and appropriately red-shifted FeK$\alpha$
line \citep[see also][]{Trinchierietal2003}.  The slope of the power law can in principle be
better constrained now because of the higher signal-to-noise at high energy.
We find however that the complex spectrum requiring a
model with several independent components  and large absorbing column
density gives too small leverage for an accurate determination of
the continuum.  We also note that the region around the FeK line is
not well modeled with the possibility of additional line components.
At lower energies the data require a  thermal plasma model with T$~$
0.6 keV and no absorption above the galactic value, in agreement with
the more tentative \chandra\ results.  The Seyfert nucleus dominates
the emission at all energies.  At soft energies (0.2-2 keV) the
unabsorbed power law and plasma components contribute equally while
above 2 keV only the nuclear power-law source is present. 

\section{Discussion} 

Owing to the large effective area of the telescope, the XMM-Newton observations
reveal new faint X-ray features in SQ that have
hitherto been missed with \chandra. In addition, improved photon statistics allows
the determination of spectral components to higher accuracy, although we still
cannot properly understand the real physical state of the gas, most likely because
it is highly inhomogeneous and possibly not in collisional equilibrium. 

XMM-Newton observations support most of the conclusions based on the 
\chandra\ observations and  add confidence to  some of the more 
tentative suggestions that were proposed.  The complexity of the X-ray 
morphology is reinforced by similarly complex spectral 
characteristics.  Nearly all \chandra\ sources are detected again by
XMM-Newton (allowing for the degraded spatial resolution) 
and we can identify additional interesting features.  Each of the 
principal emission features in the central region of SQ  will be 
briefly discussed here with an emphasis on new issues raised by the 
XMM-Newton observations. 

\begin{figure*}[th]
\epsfig{figure=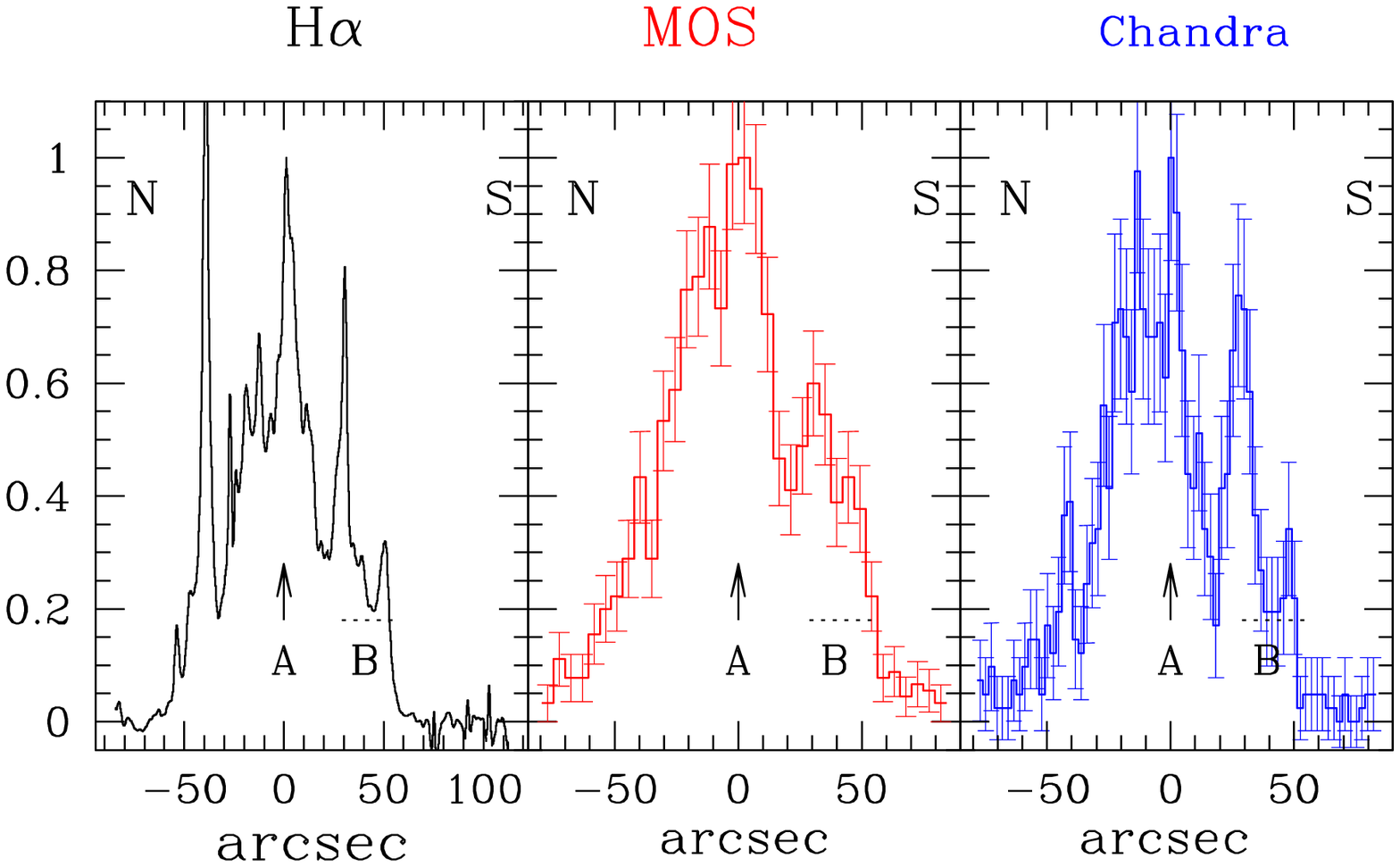,width=14.cm,clip=}
\epsfig{figure=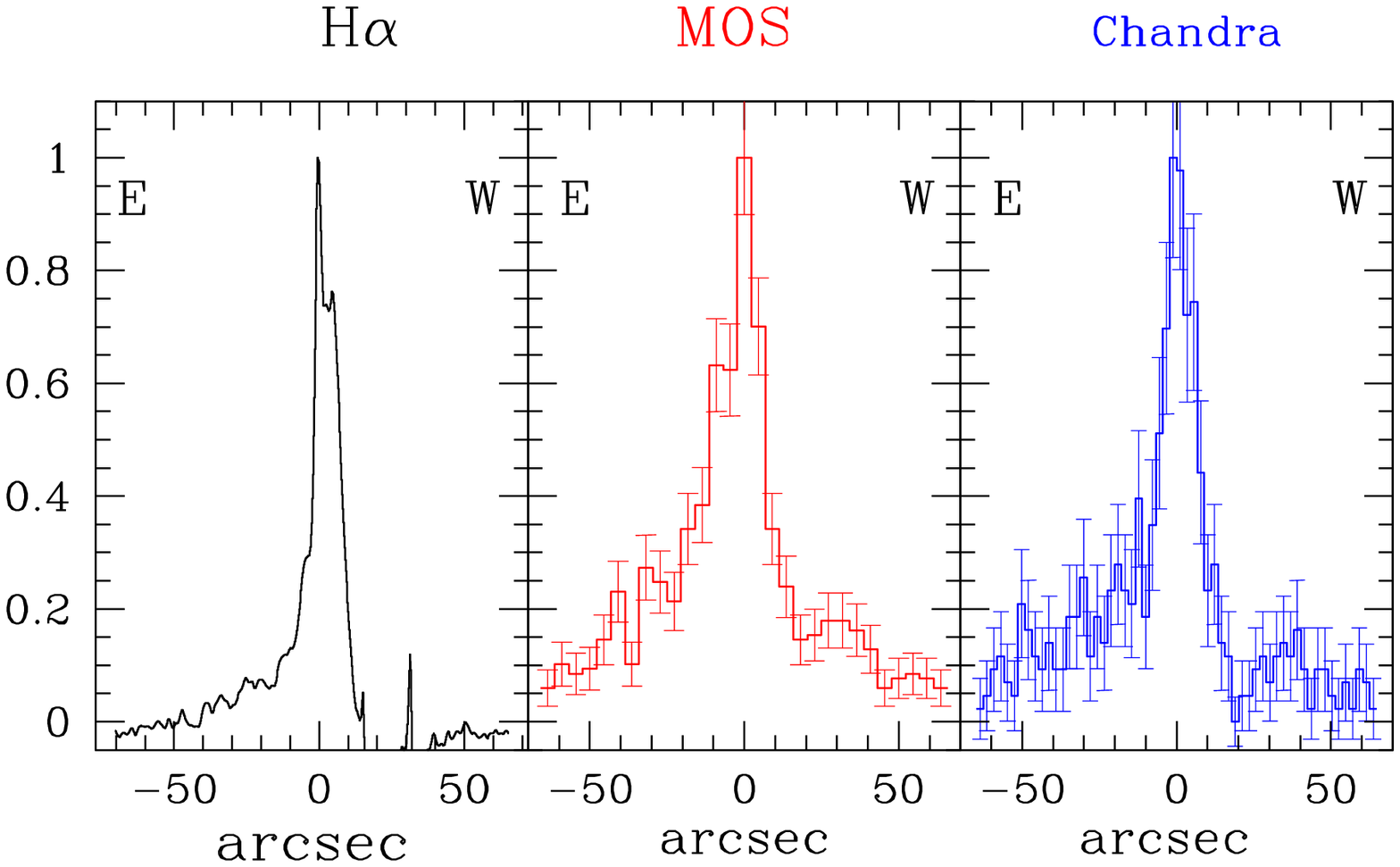,width=14.cm,clip=}
\caption{(TOP) Intensity plots of the  H$\alpha$+[N\ii] and
X-ray images (0.3-3.0 keV) in narrow cuts oriented NS (PA=2\degr)
to better compare the shapes and
distributions of X-ray and optical line emission.  (BOTTOM)
Corresponding cuts in the EW direction.  The location of
the cuts is indicated in Fig.~\ref{whichreg}.  A 15$''$ wide cut was made
in all cases.  Each step in the EPIC-MOS data corresponds to $\sim 5''$
and $\sim 2.5''$ in the \chandra\ data. All data are normalized to the
peak at 0$''$, so that shapes are directly comparable. Regions A and B
Fig.~\ref{whichreg} are identified in the top panels while  the bottom
ones are centered on A.  The negative values in the H$\alpha$ data are
due to subtraction of the NGC~7318a/b nuclei.  The higher resolution in
the \chandra\ data is apparent and distinguishes better the different
features, even at the degraded resolution used for this plot. The
XMM-Newton data provide an important confirmation on  extent of the
emission.  }
\label{xhaprof} 
\end{figure*} 

\subsection{X-ray Emission from SQ Member Galaxies 
} 
The central regions of most SQ members are now X-ray detected with
luminosities above 10$^{39}$ \ergsec. The exceptions involve the old
(NGC~7320c) and new (NGC~7318b) intruders.  NGC~7320c is far from the
current center of activity and shows no detectable \hi\ or X-ray
emitting gas.  The case of NGC~7318b, at the site of current impact, is more
complex, and might be strongly related to the SHOCK itself (see discussion in the
next section).
Remnants of the \hi\ disk of NGC~7318b are still recognizable, but no X-ray
emission clearly associated with the central regions of this galaxy is detected.
Only one of four bright emission line regions belonging to its southern 
spiral arm \citep{Sulenticetal2001} is detected (source \#2), at a luminosity of
$\sim 7 \times 10^{39}$ erg s$^{-1}$.  While high, examples of 
luminosities such as this are
now found in increasing number associated with star forming regions
\citep{ZezasGeorgantopoulosWard, Robertswarwick,
Fabbianoetal2001, Zezasfabbiano2002, 
Woltertrinchieri}.  Several models suggest that interaction/collisions could
enhance the star formation activity in galaxies \citep[e.g.][]{Jogsolomon,
Fujitanagashima, 
Bekkicouch}, so the real question becomes what distinguishes this from the other bright
regions nearby: perhaps this is simply denser/more compact or the first to evolve.  

Both intruders,
like NGC~7319 before them, are likely undergoing transformation from
late-type spiral to early-type E/S0 morphologies. Perhaps NGC~7317 and
7318a underwent a similar process in past epochs. The older optical
tidal tail with apparently associated \hi\ may be a manifestation of
that past activity. Given its ``field'' (i.e. low galaxy surface
density) environment, SQ is rich in early-type galaxies and will soon
grow even richer. This appears to be one of the characteristics of
compact groups at least as defined by the \cite{Hickson1982} catalog. The
stripping events creating early type galaxies has given rise to large
quantities of cold \citep{Williamsetal2002} and hot diffuse gas.  The
stripping events involving the intruders were apparently efficient
enough to prevent the fueling of any nuclear activity so far.
NGC~7319 is the exception and shows a ``typical'' Seyfert 2 nuclear
spectrum that can be modeled by a  double power-law with  $\Gamma \sim
1.3$ plus a heavily absorbed component.  The derived power-law slope is
flatter than  inferred from the \chandra\ data but consistent with
spectra of other Seyfert 2s \citep{DellaCeca1999, Moran2001}. 
A thermal component is required to account for excess at 
low energies.  This was also suggested by the \chandra\
data, in particular when the contribution from a circumnuclear region
of more extended emission is included in the spectrum. XMM-Newton
resolution means that any nuclear spectrum includes the circumnuclear
region (and a negligible one from the high z quasar, see \citealt{Galianni}).

\subsection{The Shock}

The sharp NS feature between NGC~7318ab and NGC~7319 shows a complex
morphology with equally intriguing spectral characteristics. Higher
resolution \chandra\ data showed several condensations and a steep W
edge to the emission. These features are confirmed by XMM-Newton albeit
with lower resolution. Comparison with data at other wavelengths
illustrates the complexity of the shock region with  shock-related
effects prominent at radio \citep{vanderhulstrots81, Williamsetal2002,
Xuetal2003}, X ray \citep{Pietschetal97, Sulenticetal2001,
Trinchierietal2003} and H$\alpha$+[N\ii].  Some of the H$\alpha$+[N\ii] emission shows a
star formation signature (both at SQ and new intruder velocities) which is
now also seen in the UV \citep{Xuetal2005}.   Examination of the high resolution Hubble
images indicate that  condensations A and B
(Fig.~\ref{whichreg}) correspond to places where the spiral arms of NGC~7318b
are visibly disrupted: in A, where one of them intersects the shock, 
one can see a clear detachment of the
inner part of this arm dominated by stellar light (emerging from the
central bulge) from the gaseous component. This can reasonably be
assumed to correspond to a region of higher gas density in the new
intruder disk. Higher resolution \chandra\ data suggest that B is 
spatially coincident with a relatively compact radio continuum feature 
\citep[see e.g. ][]{Xanthopoulosetal2004} near the south end of the shock. 
An optical spectrum \citep{Gallagheretal2001} shows the onset of shock 
induced line smearing at the same location (with new intruder velocities 
merging into SQ velocities). 

A close correspondence between X-ray and H$\alpha$+N[\ii] emission was
already noted in \cite{Trinchierietal2003} and is again evident in the
comparison with the XMM-Newton data in Fig.~\ref{hasoXima}. 
Our interference filter image
records both H$\alpha$ and [N\ii]$\lambda$6583 emission. The association with the
shock front is seen only in the  H$\alpha$+[N\ii] emission with SQ velocities
($\sim 6600$ km s$^{-1}$)\footnote{SQ and new intruder
velocity ranges are  V$>$6000 (mostly 6300-6600) km/s and 5400-6000
km/s respectively \citep{Xuetal1999, Sulenticetal2001}. 
}.
Figure \ref{hasoXima}-left shows two strong
concentrations of line  emission coincident with the Seyfert nucleus of
NGC~7319 and the shock front. There is also evidence of an extended
lower surface brightness emission component extending mostly eastward
from the shock front that we will discuss later.  The multiwavelength
morphological similarities are signatures of a close link between
different phases of the
new intruder ISM and SQ IGM.

The 1D intensity cuts along the shock front (Fig.~\ref{xhaprof}, top panels) 
show more quantitatively the
remarkable correspondence between X-ray and line emission.  The cuts are centered on A
as indicated in Fig.~\ref{whichreg}.  The total extents of the emissions
are comparable with a rather sharp boundary at $\sim 50''$ S in all
panels.  The X rays appear slightly more extended  towards the N as
more clearly shown with  the EPIC-MOS data.  A good correspondence
between line and X-ray emission is seen between $30''-50''$ to the S
while the H$\alpha$ peak at $-40''$ N, related to starburst A
\citep{Xuetal1999}, does not show a corresponding X-ray feature.
The nearby weaker X-ray peak is due to the contribution
of component C, which partially enters our cuts but is actually
displaced from the position of the H$\alpha$ maximum and extends
towards the NW.  Starburst A is coincident with an X-ray minimum as well
as an apparent \hi/CO maximum.  This probably means that the starburst
has occurred recently, with supernova activity still to come.

The steep W boundary of the X-ray shock front discovered with \chandra\ 
motivated us to suggest a NE to SW transverse component of new intruder
motion.  Its presence in the X-ray data is difficult to  understand in
the context of shocks \citep[see][]{Trinchierietal2003}.  A similarly
sharp W boundary of the optical line emission reinforces the
significance of this feature and the strong link between X-ray and
H$\alpha$+[N\ii] emitting gas.  The sharp drop is already evident in the
H$\alpha$ image (Fig.~\ref{hasoXima}) but is dramatic in the 1D intensity cuts
perpendicular to the shock front shown in Fig.~\ref{xhaprof} (bottom
panels). Several examples of sharp X-ray boundaries have now been
observed in clusters and groups \citep[i.e. ``cold
fronts",][]{Mazzottaetal,
Vikhlininetal}. However, in these, the
spatial discontinuity is thought to correspond to a significant
temperature decrease, while  SQ does not
show this effect.  Both drops in X-rays and H$\alpha$  could instead correspond to a
(pre-existing)
density discontinuity.   This in turn implies that both emissions derive from
impact with the same debris field.  
It is also conceivable that the X-ray discontinuity
corresponds to a contact discontinuity where two separate flows are
colliding. In that case, pressure and velocity would  be continuous
across the surface but the density would suddenly change.  Since
these discontinuities are generally unstable their presence might be
understood if there is a magnetic field present that could suppress
Kelvin-Helmholtz type instabilities generated by tangential flows.  The
detection of a radio continuum emission coincident with the shock front
is evidence for the presence of a magnetic field
\citep{Williamsetal2002, Xuetal2003}.  The  H$\alpha$ boundary at the
same position and with a similar drop would need to be explained in the same
context. 

\begin{figure*}
\epsfig{figure=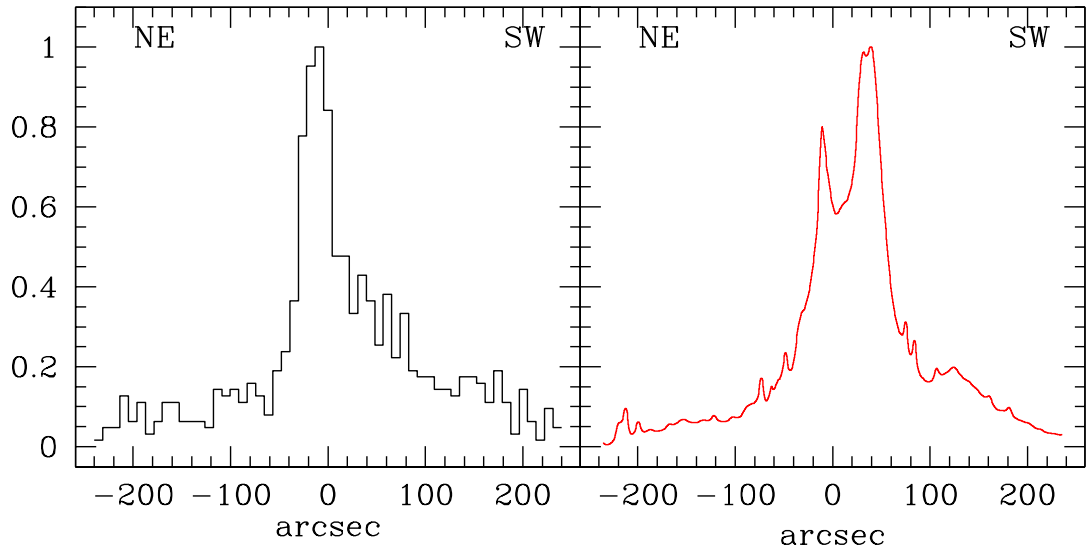,width=16cm,clip=}
\caption{Cut along the outer regions of the X-ray (left) and red
continuum (right) emission (see Fig.~\ref{whichreg}), again normalized to
their respective peaks. Each step in the
EPIC-MOS data corresponds to $\sim 7''$.   In both figures 
emission sloping to the SW is visible. 
}
\label{xmmred}
\end{figure*}

A close connection could also exist between the X-ray and dust
distributions. Xu et al. (2003) have emphasized the role of dust in the
cooling process behind the shock front resulting from the collision between
the intruder
galaxy and  SQ IGM. They have argued that it could
dominate the radiative cooling process.
Residual 60$\mu$m and
100$\mu$m emission detected with ISO
(i.e., not attributed to NGC~7319 and
the foreground NGC~7320) shows an elongated feature coincident with
the ridge observed in H$\alpha$, radio continuum and X-ray (see 
Figure 3 in Xu et al.).
\cite{Xuetal2003} have interpreted this as 
evidence for shock-related Far InfraRed emission.
They derive  a sputtering timescale for
the dust of a few times $10^6$ yr,  which is comparable to the
timescale for gas cooling due to collision with dust grains.
In their model, the expected luminosity for the dust is also 
comparable to the observed
FIR luminosity attributed to the shock region.
Since the parameters assumed in Xu et al, based on our  Chandra
observation, are basically confirmed by the present XMM-Newton data,
we do not recalculate these estimates.
However, 
to properly interpret the exact role of the dust emission in the cooling 
process, and discard other plausible interpretations (e.g. evaluate the
possible contribution from the  unshocked arm of the new intruder
along the same line of sight),
a detailed comparison between the dust distribution and the X-ray
emission in this region is required, together with a detailed treatment of the
shock physics through e.g. numerical simulations.   The complexity of the
multiphase medium that XMM-Newton data have shown will also have to be
taken into account in any detailed modeling.

The spectral properties of the shock region are not well defined but,
under the assumption of low metal abundance, imply a multi-temperature
plasma.  This is most likely the observational signature of
non-equilibrium effects.  Due to the differences in \textit{atomic
cross-sections} for ionization and recombination of different ions and
ionization stages of the same ion, their coexistence in a plasma can
mimic the signature of a multi-temperature structure.  The assumption of
equilibrium conditions that is implicit in current plasma models
applies when the ionization and recombination time scales of the plasma
are short compared to the dynamical  and cooling time scales.  This is
not the case for recently shocked gas. 
Conversely, if the plasma is expanding rapidly
due to its overpressure with respect to the ambient medium,
recombination is expected to  be delayed \citep{Breitsc94, Breitsc99}.
Unfortunately, lack of detailed knowledge of the gas dynamics and
thermal history in SQ, coupled with the limitations of the data
prevents us from applying a proper modeling to account for
non-equilibrium conditions.

Moreover we cannot discard additional effects due to a inhomogeneous
matter distribution and/or  geometrical effects.  The middle
condensation (A) with smallest N$_H$ could represent the leading edge
of the approaching shock,  already past the SQ debris field that is
visible through H$\alpha$+[N\ii] emission. The other two regions B and
C  might lag behind and  still be embedded in molecular gas responsible
for additional absorption.  The neutral hydrogen detected in these
regions is not enough to explain the inferred absorbing column
densities.  In fact no \hi\ is detected in the region covered by our
ellipse \citep{Williamsetal2002} although \hi\ features are found at
the N and S end with velocities consistent with those in the shock.  We
have interpreted the \hi\ clouds and NS shock feature as evidence for a
preexisting tidal feature possibly produced in  earlier tidal
activity \citep{Sulenticetal2001, Trinchierietal2003}.  The center of
this feature is now shocked as a result of the recent collision with the 
new intruder.  The evidence for a significant amount of debris,
traced by the H$\alpha$ emission,  implies the presence of
associated material, that could explain the higher absorption in X-rays. 

As already discussed in the context of \chandra\ results
\citep{Trinchierietal2003}, the temperatures associated with pre- and post-shock
medium are a problem in our current understanding of shocks and the physical
conditions of SQ: a bow-shock cannot be avoided if the new intruder slams 
into the debris with supersonic velocity, but the measured temperatures are lower
than what is expected from the velocity of the intruder. 
If the gas ahead of the shock is fairly hot, the 
shock Mach number is low (M=2-3), resulting in a low compression factor and 
a moderate increase of the post-shock temperature (2-4 times).   With the limitation of the current spectral analysis, we measure 
a pre-shock temperature of $\sim$  0.3-1 keV in 
``HALO" and ``TAIL" that would imply a post-shock temperature in the range
1-4 keV.  In shock, we measure
kT$\sim 1$ keV, at the lower end of the expected range (although a hint of higher
temperatures are seen in the middle blob, see Table~\ref{spectab}).  
We could of course be observing the system some time
after the collision, after significant cooling has already occurred: 
however the ``radiative" cooling times that we derive, $10^8-10^9$ yr,
are long compared to the estimated crossing time
for NGC7318b  and would further imply that the
galaxy is now at a distance of several kpc from the site of impact.  
As suggested
in \cite{Trinchierietal2003}, if the shock is oblique, most of the gas passes through the wings of the 
bow-shock under some angle and hence compression and heating are 
reduced. 
Alternatively, if the gas has been freshly 
shocked, with ionization lagging behind the entropy increase, 
the post-shock gas appears to be ``under-ionized'', i.e. it has 
essentially the same spectral characteristics as the unshocked medium.  In 
this non-equilibrium ionization scenario, compression
could be high but we could 
still derive ``low temperatures".  In this case, we have to be in a situation
where the shock is very recent, because as time goes by the system would
gradually be getting back to equilibrium conditions.  
If cooling by dust is a contributing or even dominant process, then
the timescale for the shock could be extremely short, down to 
a few $\times 10^6$ yr \citep{Xuetal2003}. In this case, if the cooling
and sputtering times are comparable,  the process could be
self-regulating and result in an almost iso-thermal shock,
explaining the lack of a temperature increase in the region.

\subsection{Separating Shocked and Diffuse X-ray Components in SQ} 

The analogies that compact groups show to rich clusters (e.g. extreme
densities with resultant strong galaxy harassment) have led to claims
that compact groups also typically show diffuse X-ray haloes. If
primordial or at least virialized they offer an important chance to map
the gravitational potential of the aggregate. Several examples of X-ray
clouds are found in compact groups: Hickson 62 represents probably the
most unambiguous example of such a component but lower luminosity and
less massive examples have  been found in other compact groups mostly
with low spiral fraction  \citep{Mulchaeyetal2003, Osmondponman2004}.
In the case of SQ, our inferences about any long-lived diffuse
component have been clouded by the strong and complex emission from the
large scale shock.  Low resolution ROSAT-PSPC and ASCA observations
were interpreted as emission dominated by a diffuse
component \citep{Sulenticetal1995, Awakietal1997}.  Subsequent
ROSAT-HRI and \chandra\ observations \citep{Pietschetal97,
Sulenticetal2001, Trinchierietal2003} showed that most of the diffuse
photons originated from the extended shock leaving little flux that
could be ascribed to diffuse emission. XMM-Newton observations reported
here provide the strongest evidence for diffuse emission in SQ because
the observed extent of the X-ray emission exceeds any reasonable estimate 
for the extent of shock related activity.

Spectral data for the extended emission also require a multi-component
plasma model in analogy with the shock region.  In fact there seems to
be a continuity in the spectral properties out to the TAIL region. The
surrounding region is apparently inhomogeneous like the shock front
with a highly variable N$_H$ column and an unrelaxed diffuse component
mixed with emission related to the ongoing shock. The complex \hi\
distribution in SQ may create or enhance the observed inhomogeneities. 
There are two X-ray components that we would like to separate: 1) shock
related emission, likely concentrated near the NS feature, where however 
emission related to the shock proper, the new intruder disk
and a more long-lived diffuse component are probably mixed, and 2) the more
long-lived diffuse component itself, of lower surface brightness and larger
distribution.   We will make use of the
narrow band H$\alpha$+[N\ii] and broad R band images to
respectively infer the extent of shock and diffuse emission
components.

The 1D EW cuts shown in Fig.~\ref{xhaprof} indicate a significant
extension towards the E in both line and X-ray emissions. The eastward
extension is more significant in the new XMM-Newton data and can be
traced up to $~60-70''$ ($\sim$40kpc) east from the shock front and
enveloping the Seyfert nucleus of NGC~7319. A similar diffuse extension
up to $\sim$ 50$''$ is observed in line emission. Emission westward of
the NS shock front is also visible in X-ray (mostly XMM-Newton) data,
but at a lower intensity (see also Fig. 5 in \cite{Trinchierietal2003}). 
H$\alpha$+[N\ii] cannot be reliably mapped here because
of problems associated with subtraction of the continuum emission near
the centers of NGC~7318ab. The simplest interpretation sees the eastern
emission extension as related to the new intruder disk.  It may be the
signature  of shocked eastern half of the disk which is reasonable
because almost no signature of neutral or warm gas is found eastward of
the shock front indicating that the disk has already passed the
front.   On the contrary, two \hi\ and numerous \hii\ regions with new
intruder velocities are observed west of the shock front. This
east-west difference suggests that either the western side of the new
intruder disk encountered much less SQ gas in its path or that it has
not yet passed through SQ. The former appears likely because the \hi\
and \hii\  associated with the western side indicate that the disk is
disrupted. The high line of sight velocity of the intruder makes this
disruption difficult to explain with a response time of t$<$10$^8$
years.   It is difficult to infer a total luminosity for the shock
related component. The bulk of the shocked emission is inferred to
follow the contours superimposed on Figure 8b where diffuse (assumed to
be forbidden [N\ii]) line emission is found.  It is the most reasonable
approach we can take given the complexity of the source and it is
reasonable to say that it is on the order of $\sim 10^{41}$
erg s$^{-1}$ which would be high for a normal spiral unless enhanced by
a collision.

Low level emission west of the shock front as well as extended emission
towards the SE (TAIL) and SW (towards NGC~7317) are the strongest
evidences for large scale diffuse X-ray emission in SQ (see
Fig.\ref{xmm+chandra} and Fig.\ref{radprof}). The irregular shape of
this component suggests that it is far from dynamically relaxed. An
optical or an X-ray diffuse halo will grow by sequential stripping in a
compact group like SQ. Earlier analysis of optical data \citep{Molesetal}
showed a halo luminosity L$\sim$L$^\ast$ and suggested a smoother
distribution of diffuse light towards the west and a very complex
distribution towards the east. The latter was interpreted as evidence
for the most recent halo building event connected with the last
passages of NGC~7320c  and stripping of NGC~7319. The halo
building component involves the optical tidal tails and associated
debris between NGC~7319  and NGC~7320c.  Star formation
condensations have also been found in this debris many tens of kpc from
group members/intruders \citep{Sulenticetal2001, Mendes2005}. Recent
ultra-deep CCD imagery \citep{Gutierrezetal} considerably expands the
size of the detected optical halo reinforcing the earlier conclusion
that it represents an even more significant luminosity, and hence
baryonic mass, component than previously estimated.

The 2D extent of the diffuse emission in both X-ray and optical light
can be inferred from Fig.~\ref{xmm+chandra} and the comparison shown in
Fig.~\ref{color} (X-ray contours on red image, left, and optical
contours on X-ray color image, right).  There is again a striking
correspondence between diffuse emissions from hot gas and the stellar
envelope in the southern half.  
Incidentally, this makes a strong case for the dynamical involvement of NGC~7317
in SQ \citep[see][]{Molesetal}. 
An extension towards NNE is
also seen in both X-ray and optical light (however, beware of the  prominent Seyfert
nucleus in the X-ray image).  This extended  emission
cannot reasonably be ascribed to the ongoing shock and was apparently
produced in earlier collisions within SQ that are hard to reconstruct now. 
The lobe-like structures are likely signatures of particular
stripping episodes that have not had time to relax.  The latest
stripping episode involving debris towards the old intruder is even
less relaxed and therefore more directly traceable to the last one or
two intrusions of NGC~7320c.  SQ is apparently an example of extensive
diffuse light in a dynamically active compact group.

Fig.~\ref{xmmred} shows a further comparison between X-ray and red
continuum emission intended to show the extent of diffuse emission
towards the SW and NE directions.  The 1D track is also shown in
Fig~\ref{whichreg} which is centered roughly SE of feature C. 
Unlike Fig.~\ref{xhaprof}, the shapes of the two
distributions are not similar.  The ``two-horned" profile in red
continuum light is due to NGC~7318 while the peak in the X-ray plot is
due to feature C and the bridge that connects it to the NS shock
front.  However diffuse emission can be traced in both plots out to at
least to 200$''$ ($~$70kpc) SW and $\sim 100''$ to the NE.  This again
indicates that the extended optical halo shows a correspondence with
the extended X-ray halo.  It is difficult to say more about the
multiphase IGM of SQ especially in the absence of 3D data.  An estimate
for the X-ray luminosity of the (non-shock) diffuse emission yields
L$_x \sim 5 \times 10^{41}$ erg s$^{-1}$  (0.5-2 keV) comparable with
the shocked component but of much lower surface brightness.  It is also
comparable to luminosities of the hot gaseous component in other low
velocity dispersion groups discussed by \cite{HelsdonPonman2005}.
SQ is a very low velocity dispersion group if the new intruder not 
included in the estimate.

\section{Conclusions}

\chandra\ and XMM-Newton have provided complimentary insights for the complex
X-ray emission from Stephan's Quintet. The strongest point source
involves the Seyfert nucleus of NGC~7319 although all members have now
been detected, except for the two intruders NGC~7320c  and NGC~7318b. 
The most unusual aspect of the X-ray emission from SQ
involves its extended emission. All of the galaxies in SQ have been
stripped by a continuing sequence of intrusions by neighboring and
likely member galaxies. The most recently stripped spiral galaxies
involve NGC~7319 and NGC~7320c. Apparently enough gas was channeled into
the nucleus of the former to stimulate an AGN while the latter is
undetected at all but optical wavelengths.  The stripping events have
given rise to a complex multiphase medium that has now been detected in
radio continuum, radio lines (\hi+CO), optical line (H$\alpha$+[N\ii]) and
X-ray. Much/most  of the radio continuum, optical line and X-ray
emission are connected with a large scale shock generated by the most
recent collision involving the ISM of NGC~7318b and the preexisting IGM
debris field from previous stripping events. We infer an extent for
the shock dominated emission in excess of D$\sim$50kpc.
The stronger emission involves a
narrow NS shock front that may reflect the existence of an extended debris
field connecting two \hi\ clouds
in the new intruder path. The more extended shock related
emission is interpreted as the shocked new intruder disk. The 2D
correspondence between the X-ray and optical line emission suggests
a common origin for these two gas phases. 

We now
detect a much more extended (D$~$130-150kpc) diffuse  component that
cannot be ascribed to the shock. This is presumably related to the
diffuse components found in other groups and clusters. No line emission
maps this X-ray structure, but rather diffuse light from a halo of
stripped stars. The complex lobe-like structure of this halo presumably
reflects a group far from a state of dynamical relaxation.

It is difficult to find an analogy to SQ
in the current X-ray literature. The closest might be the Antennae 
\citep{Fabbianoetal2003, Metzetal2004} where strong interactions 
have obviously taken place. However the analogy quickly breaks down,
because the Antennae are dominated by star formation regions and most 
of the diffuse X-ray emission is associated with these regions. In addition, 
while strongly distorted, most of the gas still resides in the galaxies. 
We do not see evidence for  widespread star formation activity in
SQ galaxies, although we see small 
star forming condensations developing in the stripped debris
\citep{Xuetal1999, Sulenticetal2001, Mendes2005}. Shocked 
emission in SQ does not follow these few features: it is
concentrated in a more localized region presumably at the sight of current impact
and appears to be more extended than traced by the H$\alpha$ emission (although
mixing with the pre-existing diffuse emission might play a role). 
The spectra of the diffuse components in SQ and the Antennae both
require multi-temperature fits, however it is very likely that this
is just the  observational signature of very complex temperature
distributions and/or non-equilibrium effects.  
In the Antennae, a detailed analysis has
shown the presence of regions at different temperatures and absorption
depths \citep{Fabbianoetal2003}.    
In SQ we both have evidence of condensations at different temperatures and expect
non-equilibrium conditions.  Unfortunately, current observations do not allow us to 
calculate the dynamics and the thermal history of the
plasma self-consistently, to properly model its physical conditions.  

\acknowledgements{ 
We thank the referee, Dr. C. Xu, for pointing out to us the relevance of
dust cooling in the shock region, and for several useful comments that
helped improve the paper. 
This research has made use of SAOImage DS9, developed by the Smithsonian 
Astrophysical Observatory.  GT acknowledges partial financial support 
by the Italian Space Agency ASI. JS acknowledge NASA support under contract
NAG5-11204. We thank Carlo Gutierrez for making his optical R-band data 
available.} 

\bibliography{biblio}

\end{document}